# Health-behaviors associated with the growing risk of adolescent suicide attempts: A data-driven cross-sectional study

Zhiyuan Wei[1], Sayanti Mukherjee[2,3]

## Abstract

**Purpose:** Identify and examine the associations between health behaviors and increased risk of adolescent suicide attempts, while controlling for socio-economic and demographic differences.
**Design:** A data-driven analysis using cross-sectional data.
**Setting:** Communities in the state of Montana from 1999 to 2017. Selected Montana as it persistently ranks among the top three vulnerable states in the U.S. over the past years.
**Subjects:** Selected 22,447 adolescents of whom 1,631 adolescents attempted suicide at least once.
**Measures:** Overall 29 variables (predictors) accounting for psychological behaviors, illegal substances consumption, daily activities at schools and demographic backgrounds, were considered.
**Analysis:** A library of machine learning algorithms along with the traditionally-used logistic regression were used to model and predict suicide attempt risk. Model performances—goodness-of-fit and predictive accuracy—were measured using accuracy, precision, recall and F-score metrics. Additionally, $\chi^2$ analysis was used to evaluate the statistical significance of each variable.
**Results:** The non-parametric Bayesian tree ensemble model outperformed all other models, with 80.0% accuracy in goodness-of-fit (F-score:0.802) and 78.2% in predictive accuracy (F-score:0.785). Key health-behaviors identified include: being sad/hopeless ($p < 0.0001$), followed by safety concerns at school ($p < 0.0001$), physical fighting ($p < 0.0001$), inhalant usage ($p < 0.0001$), illegal drugs consumption at school ($p < 0.0001$), current cigarette usage ($p < 0.0001$), and having first sex at an early age (below 15 years of age). Additionally, the minority groups (American Indian/Alaska Natives, Hispanics/Latinos) ($p < 0.0001$), and females ($p < 0.0001$) are also found to be highly vulnerable to attempting suicides.
**Conclusion:** Significant contribution of this work is understanding the key health-behaviors and health disparities that lead to higher frequency of suicide attempts among adolescents, while accounting for the non-linearity and complex interactions among the outcome and the exposure variables. Findings provide insights on key health-behaviors that can be viewed as early warning signs/precursors of suicide attempts in adolescents.

**Keywords:** mental health, health behaviors, health policy, suicide attempts among adolescents, suicide prevention, predictive analytics

[1] Graduate Student, Department of Industrial and Systems Engineering; University at Buffalo, The State University of New York
[2] Assistant Professor, Director of OASIS Laboratory, Department of Industrial and Systems Engineering, University at Buffalo, The State University of New York
[3] Corresponding Author: 411 Bell Hall, Buffalo NY 14260; Email: sayantim@buffalo.edu; Phone: 716- 645-4699

## Purpose

Suicide is ranked as the second leading cause of death among individuals aged 10-34 years in the United States.[1] Previous literature has demonstrated that suicidal behaviors among youths are associated with family-related factors[2] (e.g., family history of psychiatric disorder and suicidal acts, parental discord and disharmony, poor family communication), mental disorders[2,3] (e.g., depression, borderline or antisocial personality disorder, anxiety disorders, anorexia nervosa), negative life events[4,5] (e.g., physical/sexual abuse, bullying at schools, relationship break-ups, dating violence), health risk behaviors[6] (e.g., alcohol addiction, substance abuse) and so forth. A meta-analysis revealed suicide risk is complex in nature, and suggested future investigations should go beyond the traditional linear models (e.g., Linear/Logistic regression) to characterize its nonlinear associations.[7]

Some knowledge gaps, however, still remain in identifying the concerning health-behaviors on a daily basis that can be regarded as precursors to future suicide attempt in adolescents, and proposing the statistical model that can best capture the non-linear associations. Therefore, this paper aims to: 1) examine the confounding effects of multiple health-behaviors and their interactions on suicide attempt risk, and 2) identify the key health-behaviors (precursors) that better predict the risk of suicide attempts among U.S. adolescents, leveraging a library of advanced statistical learning models.

## Methods
### Participants

The Youth Risk Behavior Survey (YRBS) is a biennial school-based survey conducted by the Centers for Disease Control and Prevention (CDC) in collaboration with the U.S. states, territorial, and local education and health agencies to collect the self-reported data on health-related behaviors from students in grades 9-12.[8,9] In our study, we used this secondary data on health-behaviors and demographics from YRBS for the state of Montana during 1999—2017. We also collected relevant information on socioeconomic condition of the state from the Bureau of Labor Statistics corresponding to the same time-period. Here, we selected Montana because of the known fact that Montana witnesses the highest youth suicide rate in the country, persistently securing its place among the top three vulnerable states over the years.[10] However, our proposed data-driven methodology is generalized enough that can be applied to any geographical regions (contingent on data availability) to identify health-behaviors instrumental to suicide attempt risks, and predict the likelihood that an adolescent will attempt suicide.

Data collection was followed by data cleaning and variable selection in order to remove: 1) highly correlated variables (correlation coefficient $|\rho| > 0.9$) to avoid "masking effect" and aid in better interpretation of the health-behaviors and suicide attempt nexus; 2) variables with over 90% missing values; and 3) observations having missing entries. Consequently, the final dataset includes a total of 22,447 observations and 30 variables (29 predictors and the response).

### Measures

The response variable (a.k.a. dependent variable) in this study, referring to a subject's self-reported suicide attempts during the past twelve months from the instance of taking the survey, consists of two groups: "**Group-0**" denoting *not attempted suicide*; and, "**Group-1**" denoting *attempted suicide at least once*.

The predictors (a.k.a. independent variables) include a variety of individual-level information—e.g., participants' demographics (sex, age, race, education, body height and weight), illegal substances consumption (usage of alcohol, marijuana, cocaine, inhalant, steroid), assaultive behaviors (physical fighting, threatening, weapon carrying, safety concerns at school), sexual activities, risk perception (seat-belt/helmet use, riding with drunk drivers), emotional wellbeing (being sad/hopeless), and state-level socioeconomic factors (used as control variables in the analysis)—e.g., gross domestic product (GDP), per capita income and unemployment rate.

## Data Analysis

Among the 22,447 observations in the final dataset, 20,816 cases belonged to **Group-0** while 1,631 cases were classified into **Group-1**. This dataset is highly unbalanced, which can result in unreliable and biased outcomes when used for predictive models' training. Thus, we leveraged a down-sampling technique[d] to generate balanced datasets[e],[11] and conducted 30 iterations to ensure that all the observations are used at least once for model training and the model performances are robust.[12] In each iteration, the balanced dataset was randomly partitioned into training and test sets to train and test the models' performances respectively, using the 80-20% randomized holdout technique[f]. Model performance was evaluated by four statistical metrics—accuracy, precision, recall and F-score. A library of models including non-parametric machine learning models (i.e., Bagging, Random Forest, Bayesian Additive Regression Trees (BART), Support Vector Machines) as well as the traditionally-used parametric model (i.e., Logistic regression) were trained and tested (see Appendix). The models' performances were averaged across all the iterations and then, the best model was selected based on a bias-variance[g] trade-off concept[12] to ensure that the selected model outperforms other models in terms of both in-sample goodness-of-fit (i.e., modeling the relationships between predictors and risk for suicide attempts) and out-of-sample predictive accuracy (i.e., predicting the likelihood of attempting suicide conditioned on a particular situation). Finally, particular to the selected model, we calculated and ranked the independent variables' importance in predicting the response variable, and performed $\chi^2$ analysis to assess their statistical significance.

## Results

Comparing model results, we found that BART (see Appendix for model illustration) outperformed all other models, both in terms of model fit (*overall accuracy*=80.0%, *F-score*=0.802) and predictive accuracy (*overall accuracy*=78.2%, *F-score*=0.785). Note that, BART significantly performed better than the Logistic regression, establishing that the associations between risk factors (a.k.a. predictors/precursors) and suicide attempt risk (a.k.a. response variable) are nonlinear that traditional linear regression model fails to capture.

To better understand how strongly each predictor is associated with the response variable, variable inclusion proportions[h] (VIP) was computed for each predictor. Larger value of VIP indicates higher importance of the variables in explaining and predicting the response variable.

---

[d] This method randomly selects the equal number of observations (sampling without replacement) from Group-0 to match with the number of observations in Group-1.
[e] Balanced dataset consists of equal-sized samples from each group Group-0 and Group-1.
[f] This technique indicates 80% of the data is randomly selected as "training set" for training the models and the remaining 20% is held out as "test set" for testing the model's predictive performance.
[g] Bias-variance tradeoff is a widely used technique in machine learning to minimize the generalization error by balancing the error measured by bias and variance.
[h] VIP indicates the fraction of times a given predictor is used in growing a classification tree, and can be used to measure variable importance.

[13,14] The ranking of all the 29 predictors, based on their VIPs from high to low, is exhibited in **Fig. 1**.

For the sake of brevity, this paper further discussed the top ten predictors (from **Fig. 1**) associated with adolescent suicide attempt risk. The statistical significance of the top ten predictors are shown in **Tab. 1**, where the predictors are categorized into two groups: health-related behaviors (seven variables) and demographic characteristics (three variables). Note that, all the top ten predictors are categorical with different levels (values)[i], and are statistically significant ($p < 0.0001$). **Tab. 1** also provides detailed information on distribution of each predictor across different levels with respect to the response variable.

The seven most important health-related behaviors associated with higher suicide attempt risk in adolescents are shown in **Tab. 1** with VIPs from high to low. The foremost predictor *being sad or hopeless*, reported whether a subject had felt sad/hopeless persistently for almost every day over two or more weeks in the past year, during which they stopped doing their usual activities. Our analysis revealed that 78% of the subjects who attempted suicides (**Group-1**) were suffering from sadness or hopelessness compared to 23% of those who never attempted suicide (**Group-0**), highlighting that mental wellbeing is highly correlated with suicidal behaviors. *Safety concerns at school*, indicating number of days a subject was absent from school due to unsafe feeling at school during the past 30 days, is the second most important predictor, showing that 15% of Group-1 subjects was absent from school for atleast a day, compared to 3% of Group-0 subjects. We also found *physical fighting at schools* to be another important precursor—48.8% of Group-1 subjects were involved with physical fighting at school atleast once, compared to only 22% of Group-0 subjects. Similarly, other important health-related behaviors in **Tab. 1** are inhalant usage, illegal drugs consumption at school, current cigarette usage, and having first sex at an early age (below 15 years of age).

The three most significant demographic variables in **Tab. 1** are sex, race and education (in descending order by VIP). Our results indicated that females are more vulnerable to attempting suicide than males—70% of Group-1 subject were females, compared to 52% female subjects in Group-0. Minority groups (e.g., American Indian and Alaska native; Hispanic/Latino) are also found to be more vulnerable compared to their White or Asian peers. For example, 10.5% of Group-1 subjects were American Indian and Alaska native compared to their Group-0 counterparts (4.7%). Similarly, 8.3% of Group-1 subjects were Hispanic/Latino compared to their Group-0 counterparts (4.9%). Adolescents in 9th and 10th grades are found to be at a higher risk of suicide attempt than those in 11th and 12th grades.

## Discussion
### Summary

The major findings from this study indicate that certain health-related behaviors among adolescents have strong correlations with the risk of suicide attempts. Being sad or hopeless is the most important predictor of adolescent suicide attempt risk. Up to 78% adolescents who attempted suicide had the persistent feelings of sadness and hopelessness that hindered them from doing usual activities. Similar findings were reported in a study that targeted the group of South African secondary school learners.[15] Other health-related behaviors, mostly observed in a school environment such as absence from school due to safety concerns, involving in physical fighting, illegal drugs usage in school property, also play critical roles in predicting the risk of suicide attempt. For instance, those who were absent from school over six days had a 50% likelihood of attempting suicide. These school-based concerning behaviors can be easily tracked on a daily basis, and could be regarded as early warning signs of suicide attempt risk. Higher frequency of addictive substance consumption (e.g., inhalant, cigarette) can also reflect a higher risk of attempting suicides among youth. Engaging in sexual activity during early adolescent period is another key predictor that has shown significant associations with increased risk of suicide attempt.

---

[i] The categorical variable contains different levels, where each level can be treated as a subgroup. For example, the variable "sex" includes two levels of information: female and male. That is to say, variable "sex" can separate the whole dataset into two subgroups: being female or male.

Besides health-related behaviors, this study establishes that the risk of suicide attempt significantly differs by sex, race and education. Health disparities exist where certain demographic subgroups are relatively vulnerable to attempting suicide, such as adolescent females, American Indian and Alaska native and Hispanic/Latino adolescents, and youths studying in 9$^{th}$ and 10$^{th}$ grades.

## Limitations

Health-behaviors identified in this study illustrate the correlations/associations, but not necessarily the causality of suicide attempt risk. More information related to adolescents' family background, pre-existing mental/physical health conditions, school GPA, and/or social determinants of health (e.g., social norms and attitudes) can help to better understand the causality of suicide attempt risk in adolescents.

## Significance

Unlike previous studies, this study identified, examined and ranked a multitude of health-related behaviors and demographic characteristics of adolescents associated with suicide attempt risks while capturing their nonlinear and complex interactions, leveraging an advanced data-driven approach. Different from traditionally-used linear models, our proposed nonlinear model (BART) demonstrates higher accuracy in predicting suicide attempt risk and provides a variable importance ranking indicating how strongly they are correlated with the response. We found that a majority of adolescents attempting suicides had persistent feelings of sadness and/or hopelessness for over two weeks. This study also established that certain school-based behaviors are key precursors to suicide attempt risks. In this view, identifying the precursors by the adolescents' family and school staffs, and monitoring the mental/physical wellbeing of students will help to develop informed suicide prevention strategies, minimizing the overall suicide risk. Higher frequency of addictive substance usage (inhalant, cigarette) is also found to be a key factor of the growing risk of suicide attempt; this indicates that local government and communities might need to impose a stricter surveillance system to restrain addictive substances sales to youth.

# SO WHAT?

## *What is already known on this topic?*

Previous studies modeled the associations of adolescent suicide attempts with a single or a few selected health-behaviors assuming a linear form.

## *What does this article add?*

Our study established that the associations between health-behaviors and suicide attempt risk is nonlinear and can be best captured by nonparametric ensemble tree-based model. This study also identified and ranked the key health-behaviors that can be viewed as precursors / early warning signs for future suicide attempts. Factors include persistent feelings of sadness and/or hopelessness, school absenteeism, physical fighting, illegal drug usage, under-aged sexual activities.

### *What are the implications for health promotion practice or research?*

The identified key health-behaviors and their ranking could help stakeholders—e.g., adolescents' families, school staffs, school district/boards and governments/community leaders to make informed suicide prevention strategies and promote mental health among adolescents. Strategies may range from monitoring concerning/risky behaviors within the school environment, regulating drugs sale at the community level, and address the health inequity issues among vulnerable population such as females and minority racial groups.  This study paves a new path for predicting future suicide attempt risk leveraging advanced data-driven machine learning models.

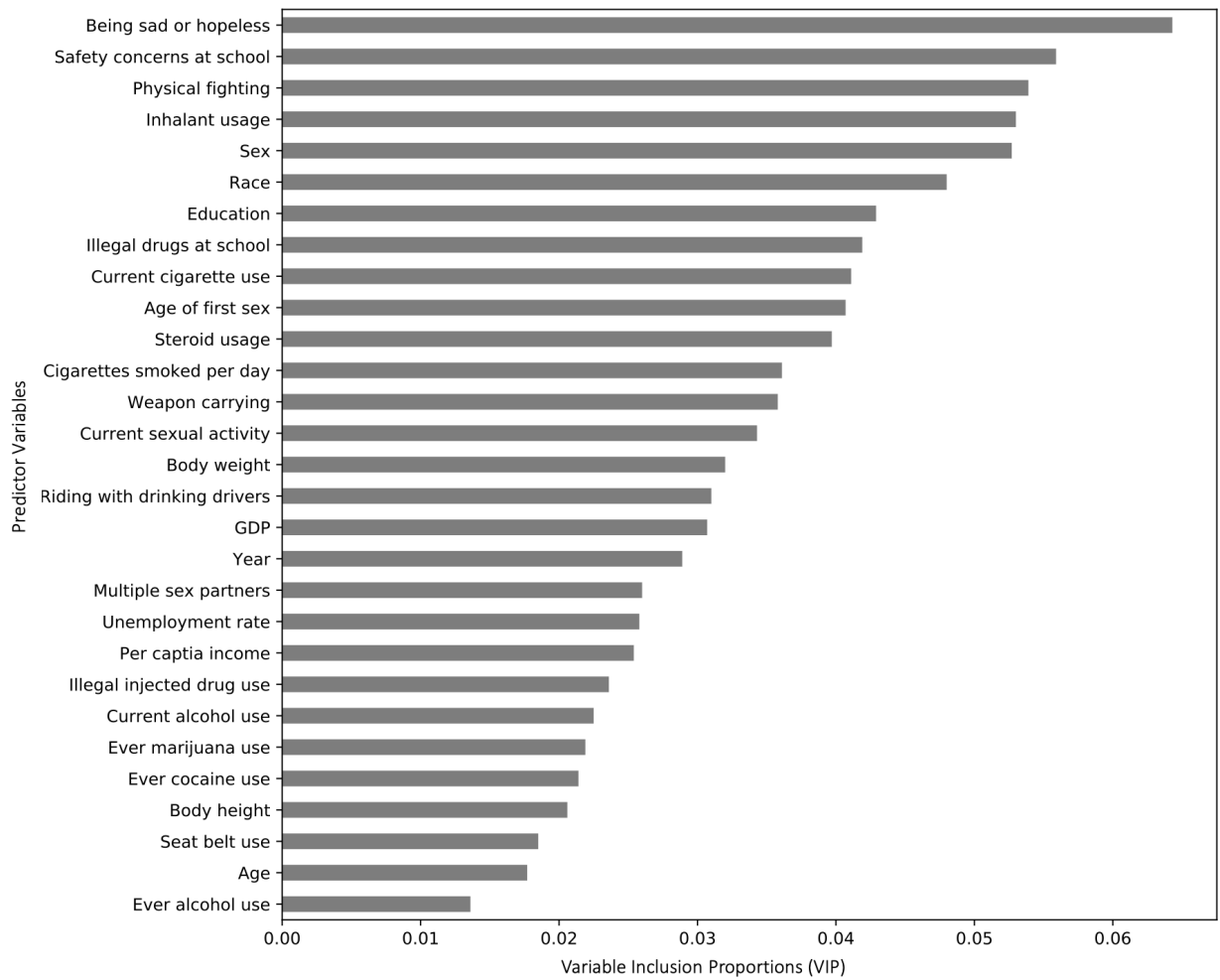

**Figure 1.** Variable importance ranking of all 29 predictors, including health behaviors and socioeconomic and demographics factors, as measured by variable inclusion proportions

**Table 1.** Top ten factors (seven health-related behaviors and three demographic factors) associated with the adolescent suicide attempt risk

| Seven Health-related Behaviors | Group 0 (N=20,816) | Group 1 (N=1,631) | $\chi^2$ Statistics | P-Value |
|---|---|---|---|---|
| **Being sad or hopeless (VIP=0.064)** | | | | |
| Yes | 4751 (23%) | 1277 (78%) | 2366.5 | <.0001*** |
| No | 16065 (77%) | 354 (22%) | | |
| **Safety concerns at school (VIP=0.056)** | | | | |
| 0 days | 20199 (97%) | 1391 (85%) | | |
| 1 day | 411 (2%) | 109 (7%) | | |
| 2-3 days | 134 (0.7%) | 70 (4.3%) | 687.13 | <.0001*** |
| 4-5 days | 27 (0.1%) | 17 (1%) | | |
| ≥ 6 days | 45 (0.2%) | 44 (2.7%) | | |
| **Physical fighting (VIP=0.054)** | | | | |
| 0 times | 16177 (78%) | 835 (51.2%) | | |
| 1 time | 2341 (11%) | 285 (17.5%) | | |
| 2-3 times | 1506 (7.2%) | 269 (16.5%) | | |
| 4-5 times | 321 (1.5%) | 92 (5.6%) | 761.13 | <.0001*** |
| 6-7 times | 154 (0.7%) | 41 (2.5%) | | |
| 8-9 times | 74 (0.4%) | 22 (1.4%) | | |
| 10-11 times | 25 (0.1%) | 10 (0.6%) | | |
| ≥ 12 times | 218 (1.1%) | 77 (4.7%) | | |
| **Inhalant usage (VIP=0.053)** | | | | |
| 0 times | 18945 (91%) | 1113 (68.3%) | | |
| 1-2 times | 1072 (5.15%) | 213 (13%) | | |
| 3-9 times | 477 (2.3%) | 119 (7.3%) | 1088.8 | <.0001*** |
| 10-19 times | 167 (0.8%) | 74 (4.5%) | | |
| 20-39 times | 72 (0.35%) | 36 (2.2%) | | |
| ≥ 40 times | 83 (0.4%) | 76 (4.7%) | | |
| **Illegal drugs at school (VIP=0.042)** | | | | |
| Yes | 4296 (20%) | 667 (41%) | 359.22 | <.0001*** |
| No | 16520 (80%) | 964 (59%) | | |
| **Current cigarette uses (VIP=0.041)** | | | | |
| 0 days | 17744 (85.2%) | 925 (56.7%) | | |
| 1-2 days | 897 (4.3%) | 151 (9.3%) | | |
| 3-5 days | 443 (2.1%) | 69 (4.2%) | | |
| 6-9 days | 249 (1.2%) | 74 (4.5%) | 999.36 | <.0001*** |
| 10-19 days | 330 (1.6%) | 60 (3.7%) | | |
| 20-29 days | 386 (1.9%) | 102 (6.3%) | | |
| All 30 days | 767 (3.7%) | 250 (15.3%) | | |
| **Age of first sex (VIP=0.040)** | | | | |
| Never had sex | 12505 (60.1%) | 500 (30.7%) | | |
| 11 years old or younger | 288 (1.4%) | 134 (8.2%) | | |
| 12 years old | 277 (1.3%) | 74 (4.5%) | 1081.5 | <.0001*** |
| 13 years old | 742 (3.6%) | 172 (10.6%) | | |
| 14 years old | 1671 (8%) | 277 (17%) | | |
| 15 years old | 2517 (12.1%) | 299 (18.3%) | | |
| **Three Demographics** | **Group 0 (N=20,816)** | **Group 1 (N=1,631)** | **$\chi^2$ Statistics** | **P-Value** |
| **Sex (VIP=0.053)** | | | | |
| Female | 10778 (52%) | 1151 (70%) | 213.76 | <.0001*** |
| Male | 10038 (48%) | 480 (30%) | | |
| **Race (VIP=0.048)** | | | | |
| American Indian/Alaska Native | 984 (4.7%) | 171 (10.5%) | | |
| Asian | 180 (0.9%) | 16 (1.0%) | | |
| Black or African American | 132 (0.6%) | 14 (0.8%) | 174.89 | <.0001*** |
| Hispanic/Latino | 1013 (4.9%) | 136 (8.3%) | | |
| Native Hawaiian/Pacific Islander | 81 (0.4%) | 13 (0.8%) | | |
| White | 17595 (84.5%) | 1189 (73.0%) | | |
| **Education (VIP=0.042)** | | | | |
| 9th grade | 5368 (26%) | 508 (31%) | | |
| 10th grade | 5601 (27%) | 489 (30%) | 67.936 | <.0001*** |
| 11th grade | 5176 (25%) | 401 (25%) | | |
| 12th grade | 4671 (22%) | 233 (14%) | | |

# References

1. Hedegaard H, Curtin SC, Warner M. Suicide mortality in the United States, 1999--2017. 2018.
2. Bilsen J. Suicide and youth: risk factors. *Front psychiatry*. 2018;9:540.
3. Bae S, Ye R, Chen S, Rivers PA, Singh KP. Risky behaviors and factors associated with suicide attempt in adolescents. *Arch suicide Res*. 2005;9(2):193-202.
4. Hepburn L, Azrael D, Molnar B, Miller M. Bullying and suicidal behaviors among urban high school youth. *J Adolesc Heal*. 2012;51(1):93-95.
5. Swahn MH, Simon TR, Hertz MF, et al. Linking dating violence, peer violence, and suicidal behaviors among high-risk youth. *Am J Prev Med*. 2008;34(1):30-38.
6. Beautrais AL. Risk factors for suicide and attempted suicide among young people. *Aust New Zeal J Psychiatry*. 2000;34(3):420-436.
7. Franklin JC, Ribeiro JD, Fox KR, et al. Risk factors for suicidal thoughts and behaviors: a meta-analysis of 50 years of research. *Psychol Bull*. 2017;143(2):187.
8. CDC. 2017 YRBS national, state, and district combined datasets user's guide. 2017.
9. Kann L, McManus T, Harris WA, et al. Youth risk behavior surveillance—United States, 2017. *MMWR Surveill Summ*. 2018;67(8):1.
10. Herling D. YOUTH SUICIDE IN MONTANA. *Mont Bus Q*. 2019;57(2):6-9.
11. Provost F. Machine learning from imbalanced data sets 101. In: *Proceedings of the AAAI 2000 Workshop on Imbalanced Data Sets*. Vol 68. ; 2000:1-3.
12. James G, Witten D, Hastie T, Tibshirani R. *An Introduction to Statistical Learning*. Vol 112. Springer; 2013.
13. Chipman HA, George EI, McCulloch RE, others. BART: Bayesian additive regression trees. *Ann Appl Stat*. 2010;4(1):266-298.
14. Kapelner A, Bleich J. bartMachine: Machine learning with Bayesian additive regression trees. *arXiv Prepr arXiv13122171*. 2013.
15. James S, Reddy SP, Ellahebokus A, Sewpaul R, Naidoo P. The association between adolescent risk behaviours and feelings of sadness or hopelessness: a cross-sectional survey of south African secondary school learners. *Psychol Health Med*. 2017;22(7):778-789.